\documentclass[pre,twocolumn,amsmath,amssymb,floatfix]{revtex4-1}

\usepackage{graphicx}
\usepackage{color}

\begin{document}

\title{Analytically solvable autocorrelation function for weakly correlated interevent times} 
\author{Hang-Hyun Jo}
\email{hang-hyun.jo@apctp.org}
\affiliation{Asia Pacific Center for Theoretical Physics, Pohang 37673, Republic of Korea}
\affiliation{Department of Physics, Pohang University of Science and Technology, Pohang 37673, Republic of Korea}
\affiliation{Department of Computer Science, Aalto University, Espoo FI-00076, Finland}

\date{\today}

\begin{abstract}
    Long-term temporal correlations observed in event sequences of natural and social phenomena have been characterized by algebraically decaying autocorrelation functions. Such temporal correlations can be understood not only by heterogeneous interevent times (IETs) but also by correlations between IETs. In contrast to the role of heterogeneous IETs on the autocorrelation function, yet little is known about the effects due to the correlations between IETs. In order to rigorously study these effects, we derive an analytical form of the autocorrelation function for the arbitrary IET distribution in the case with weakly correlated IETs, where the Farlie-Gumbel-Morgenstern copula is adopted for modeling the joint probability distribution function of two consecutive IETs. Our analytical results are confirmed by numerical simulations for exponential and power-law IET distributions. For the power-law case, we find a tendency of the steeper decay of the autocorrelation function for the stronger correlation between IETs. Our analytical approach enables us to better understand long-term temporal correlations induced by the correlations between IETs.
\end{abstract}

\maketitle

\section{Introduction}\label{sec:intro}

A variety of dynamical processes in natural and social phenomena have been described by a series of events or event sequences showing non-Poissonian or bursty nature. Examples include solar flares~\cite{Wheatland1998WaitingTime}, earthquakes~\cite{Corral2004LongTerm, deArcangelis2006Universality}, neuronal firings~\cite{Kemuriyama2010Powerlaw}, and human activities~\cite{Barabasi2005Origin, Karsai2018Bursty}. Temporal correlations in such bursty event sequences have often been characterized in terms of autocorrelation functions~\cite{Kantelhardt2001Detecting, Allegrini2009Spontaneous, Karsai2012Universal, Yasseri2012Dynamics}. The autocorrelation function for an event sequence $x(t)$ is defined with delay time $t_d$ as follows:
\begin{equation}
\label{eq:auto_define}
  A(t_d)\equiv \frac{ \langle x(t)x(t+t_d)\rangle_t- \langle x(t)\rangle^2_t}{ \langle x(t)^2\rangle_t- \langle x(t)\rangle^2_t},
\end{equation}
where $\langle\cdot\rangle_t$ denotes a time average. The event sequence $x(t)$ can be considered to have the value of $1$ at the moment of event occurred, $0$ otherwise. For the event sequences with long-term memory effects, one typically finds an algebraically decaying behavior with a decaying exponent $\gamma$:
\begin{equation}
    A(t_d)\sim t_d^{-\gamma}. 
\end{equation}
The decaying exponent $\gamma$ is known to be related to other exponents characterizing temporal correlations, such as Hurst exponent $H$~\cite{Peng1994Mosaic} and the scaling exponent of the power spectral density $\eta$~\cite{Bak1987Selforganized, Weissman1988Noise, Ward2007Noise}, via the relations $H=1-\gamma/2$ and $\eta=1-\gamma$~\cite{Kantelhardt2001Detecting, Allegrini2009Spontaneous, Rybski2009Scaling, Rybski2012Communication}. Temporal correlations measured by $A(t_d)$ can be fully understood not only by heterogeneous properties of time intervals between two consecutive events, i.e., interevent times (IETs), but also by correlations between IETs~\cite{Jo2017Modeling}. 

The heterogeneities of IETs, denoted by $\tau$, indicate the presence of multiple timescales or even the absence of characteristic timescales (i.e., scale-free), which is often related to nonhomogeneous or time-dependent Poisson processes~\cite{deArcangelis2016Statistical}. Many empirical analyses~\cite{Karsai2018Bursty} have shown that heterogeneities of IETs can be characterized by heavy-tailed or power-law IET distributions $P(\tau)$ with a power-law exponent $\alpha$:
\begin{equation}
    \label{eq:Ptau_simple}
    P(\tau)\sim \tau^{-\alpha},
\end{equation}
which readily implies clustered short IETs even without correlations between IETs. This phenomenon has been called bursts, namely, rapidly occurring events within short time periods alternating with long inactive periods~\cite{Barabasi2005Origin, Karsai2018Bursty}. It has been known that bursty interactions between individuals have a strong influence on the dynamical processes taking place in a network of individuals~\cite{Vazquez2007Impact, Karsai2011Small, Miritello2011Dynamical, Rocha2011Simulated, Jo2014Analytically, Delvenne2015Diffusion, Artime2017Dynamics, Hiraoka2018Correlated}. When IETs are fully uncorrelated with each other as in renewal processes~\cite{Mainardi2007Poisson}, the scaling relations between $\alpha$ and $\gamma$ have been analytically derived as~\cite{Lowen1993Fractal}
\begin{eqnarray}
    \label{eq:alpha_gamma}
    \begin{tabular}{ll}
        $\alpha+\gamma=2$ & for $1<\alpha\leq 2$,\\
        $\alpha-\gamma=2$ & for $2<\alpha\leq 3$.
    \end{tabular}
\end{eqnarray}
This implies that the decaying behavior of the autocorrelation function can be accounted for solely by the power-law tail of the IET distribution. 

In contrast to the role of heterogeneous IETs on the long-term temporal correlations, the effects due to the correlations between IETs are far from being fully explored, except for a few recent works: These effects were studied, e.g., by comparing the original, empirical autocorrelation functions to those calculated for the randomized event sequences~\cite{Karsai2012Universal, Rybski2012Communication}. In other works, modeling and numerical approaches were taken for investigating how strong correlations between IETs should be present to violate the scaling relations in Eq.~\eqref{eq:alpha_gamma}~\cite{Vajna2013Modelling, Jo2017Modeling, Lee2018Hierarchical}. This situation clearly calls for a rigorous, analytical approach to the role of correlations between IETs in temporal correlations. For this, the correlations between IETs can be quantified by a memory coefficient $M$~\cite{Goh2008Burstiness} among others such as local variation~\cite{Shinomoto2003Differences} or bursty trains~\cite{Karsai2012Universal}. The memory coefficient is defined as the Pearson correlation coefficient between two consecutive IETs, whose value for a sequence of $n$ IETs, i.e., $\{\tau_1,\cdots, \tau_n\}$, can be estimated by
\begin{equation}
    M \equiv\frac{1}{n - 1}\sum_{i=1}^{n-1}\frac{(\tau_i - \mu_1)(\tau_{i+1} - \mu_2)}{\sigma_1 \sigma_2},
    \label{eq:memory_original}
\end{equation}
where $\mu_1$ ($\mu_2$) and $\sigma_1$ ($\sigma_2$) are the average and the standard deviation of the first (last) $n-1$ IETs, respectively. Positive $M$ implies that the large (small) IETs tend to be followed by large (small) IETs. Negative $M$ indicates the opposite tendency, while $M=0$ means the uncorrelated IETs. We mainly focus on the case with $M\geq 0$, based on the empirical observations~\cite{Goh2008Burstiness, Wang2015Temporal, Guo2017Bounds, Bottcher2017Temporal}.

In order to rigorously study the effects of correlations between IETs on the autocorrelation function, we derive an analytical form of the autocorrelation function for the arbitrary $P(\tau)$ and for small $M$, i.e., in the case with weakly correlated IETs, where the Farlie-Gumbel-Morgenstern copula~\cite{Nelsen2006Introduction, Takeuchi2010Constructing} is adopted for modeling the joint probability distribution function of two consecutive IETs. Our analytical results are numerically confirmed for both exponential and power-law IET distributions. In particular, for the power-law case, we find the steeper decay of the autocorrelation function for the stronger correlation between IETs: The apparent decaying exponent $\gamma$ is found to increase with $M$. Our finding can help us to understand the effects of correlations between IETs on other measures such as Hurst exponent $H$ and the scaling exponent of the power spectral density $\eta$ because $\gamma$, $H$, and $\eta$ are not independent of each other as mentioned.

\section{Results}\label{sec:result}

\subsection{Analysis}\label{subsec:analysis}

We analyze the autocorrelation function $A(t_d)$ in Eq.~\eqref{eq:auto_define}. Since $A(t_d=0)=1$ is obvious, we consider the case with $t_d>0$ unless otherwise stated. Note that for a Poisson process, i.e., without any memory effects in it, $A(t_d)=0$ for all $t_d>0$. The event sequence $x(t)$ has the value of $1$ at the moment of event occurred, $0$ otherwise. Each event is assumed to have a duration of $1$. Since events may overlap with each other due to their duration, we set the lower bound of IETs as $1$, i.e., $\tau_{\rm min}=1$, for the sake of simplicity. Then for an event sequence with $n$ events during the time period $T$, we get $\lambda \equiv \langle x(t)\rangle_t = n/T = 1/\mu$, with $\mu$ denoting the mean IET. Using this $\lambda$, one can write
\begin{equation}
	\langle x(t)x(t+t_d)\rangle_t = \lambda \sum_{k=1}^\infty P_k(t_d),
\end{equation}
where $P_k(t_d)$ is the probability that two events occurred in times $t$ and $t+t_d$ are separated by exactly $k$ interevent times (IETs) for $k=1,2,\cdots$. Using the joint probability distribution function (PDF) of $k$ consecutive IETs, denoted by $P(\tau_1,\cdots,\tau_k)$, one gets 
\begin{equation}
	\label{eq:Pk_td}
	P_k(t_d) = \prod_{i=1}^k\int_0^\infty d\tau_i P(\tau_1,\cdots,\tau_k)\ \delta\left(t_d-\sum_{i=1}^k \tau_i\right),
\end{equation}
where $\delta(\cdot)$ is a Dirac delta function. Then the autocorrelation function in Eq.~\eqref{eq:auto_define} can be rewritten as
\begin{equation}
	\label{eq:Atd}
	A(t_d)=\frac{ \sum_{k=1}^\infty P_k(t_d) - \lambda}{ 1-\lambda },
\end{equation}
where we have used $\langle x(t)^2\rangle_t= \langle x(t)\rangle_t= \lambda$ as $x(t)=0,1$. 

Since we only consider the correlations between two consecutive IETs, $P(\tau_1,\cdots,\tau_k)$ in Eq.~\eqref{eq:Pk_td} can be factorized in terms of joint PDFs of two consecutive IETs, i.e., $P(\tau_i,\tau_{i+1})$ for $i=1,\cdots,n-1$. Precisely, by assuming that an IET, $\tau_{i+1}$, is conditioned only by its previous IET, $\tau_i$, namely,
\begin{equation}
    P(\tau_{i+1}|\tau_i,\tau_{i-1},\cdots)= P(\tau_{i+1}|\tau_i),
\end{equation}
one obtains
\begin{equation}
	P(\tau_1,\cdots,\tau_k) = \prod_{i=1}^{k-1} P(\tau_i, \tau_{i+1}) \Big/ \prod_{i=2}^{k-1} P(\tau_i).
    \label{eq:Ptau1tauk}
\end{equation}

For modeling $P(\tau_i,\tau_{i+1})$, we adopt a Farlie-Gumbel-Morgenstern (FGM) copula among others~\cite{Nelsen2006Introduction, Takeuchi2010Constructing} because the FGM copula is simple and analytically tractable, despite the range of correlation being somewhat limited, which will be discussed later. The FGM copula is originally defined as a function $C$ joining a bivariate cumulative distribution function (CDF) to their one-dimensional marginal CDFs such that 
\begin{eqnarray} 
  G(x_1,x_2) &=& C[u_1(x_1),u_2(x_2)] \nonumber\\
  &=& u_1u_2[1+r(1-u_1)(1-u_2)],
\end{eqnarray}
where $u_1$ ($u_2$) is a CDF of variable $x_1$ ($x_2$), and $r$ controls the correlation between $x_1$ and $x_2$~\cite{Takeuchi2010Constructing, Nelsen2006Introduction}. The bivariate PDF of $x_1$ and $x_2$ is obtained by 
\begin{equation} 
  \frac{\partial^2G(x_1,x_2)}{\partial x_1\partial x_2}=P_1(x_1)P_2(x_2)[1+r(2u_1-1)(2u_2-1)], 
\end{equation}
where $P_1(x_1)$ and $P_2(x_2)$ denote PDFs of $x_1$ and $x_2$, respectively. This FGM copula has been applied, e.g., for modeling the bivariate luminosity function of galaxies~\cite{Takeuchi2010Constructing} and for the health care data analysis~\cite{Prieger2002Flexible}. 

The joint PDF of two consecutive IETs based on the FGM copula is written as 
\begin{equation}
	P(\tau_i,\tau_{i+1}) = P(\tau_i) P(\tau_{i+1}) \left[1+r f(\tau_i)f(\tau_{i+1})\right],
	\label{eq:Ptautau}
\end{equation}
where 
\begin{equation}
    f(\tau) \equiv 2F(\tau)-1,\ F(\tau) \equiv \int_0^{\tau} d\tau' P(\tau').
\end{equation}
Here $P(\tau_i)$ and $P(\tau_{i+1})$ are assumed to have the same functional form. The range of the parameter $r$ is given as $|r|\leq 1$ because $|f(\tau)|\leq 1$ from $0\leq F(\tau)\leq 1$ and $P(\tau_i,\tau_{i+1})\geq 0$. To relate $r$ to $M$ in Eq.~\eqref{eq:memory_original}, we redefine $M$ as
\begin{equation}
    M \equiv \frac{ \langle \tau_i\tau_{i+1}\rangle - \mu^2 }{\sigma^2 },
    \label{eq:M_new}
\end{equation}
where 
\begin{equation}
    \langle \tau_i\tau_{i+1} \rangle \equiv \int_0^\infty d\tau_i \int_0^\infty d\tau_{i+1} \tau_i\tau_{i+1} P(\tau_i,\tau_{i+1}),
\end{equation}
and $\mu$ and $\sigma$ are the mean and standard deviation of IETs, respectively. Using Eq.~\eqref{eq:Ptautau} we get 
\begin{equation}
    M =\frac{r}{\sigma^2} \left[ \int_0^\infty d\tau \tau P(\tau)f(\tau) \right]^2 \equiv ar.
    \label{eq:Mr_ratio}
\end{equation}
The ratio $a$ between $M$ and $r$ is determined only by $P(\tau)$, irrespective of the correlations between IETs. Note that the upper bound of $a$ is $1/3$ for any $P(\tau)$, hence $|M|\leq 1/3$~\cite{Schucany1978Correlation}. Due to this bound, applications of the FGM copula are limited to weakly correlated cases.

By plugging Eq.~\eqref{eq:Ptautau} into Eq.~\eqref{eq:Ptau1tauk}, we get
\begin{equation}
	P(\tau_1,\cdots,\tau_k) = \prod_{i=1}^k P(\tau_i) \prod_{i=1}^{k-1} \left[1+r f(\tau_i) f(\tau_{i+1})\right]. 
    \label{eq:Ptau1tauk_expand}
\end{equation}
As it is not straightforward to analyze Eq.~\eqref{eq:Ptau1tauk_expand}, we focus on the weakly correlated case with $|r|\ll 1$. In this range of $r$, one can expand Eq.~\eqref{eq:Ptau1tauk_expand} up to the first order of $r$ as follows:
\begin{eqnarray}
    &&P(\tau_1,\cdots,\tau_k) \approx \prod_{i=1}^k P(\tau_i) \left[1+ r \sum_{i=1}^{k-1} f(\tau_i)f(\tau_{i+1}) + \mathcal{O}(r^2)\right], \nonumber \\
\end{eqnarray}
which enables us to calculate the Laplace transform of $P_k(t_d)$ in Eq.~\eqref{eq:Pk_td}: 
\begin{eqnarray}
    \widetilde{P_k}(s) &\equiv& \int_0^\infty dt_d P_k(t_d) e^{-st_d}\nonumber\\
    &\approx& \tilde{P}(s)^k + r(k-1)\tilde{P}(s)^{k-2}\tilde{Q}(s)^2 + \mathcal{O}(r^2),
\end{eqnarray}
with
\begin{eqnarray}
	\tilde{P}(s) &\equiv& \int_0^\infty d\tau P(\tau)e^{-s\tau},\\
	\tilde{Q}(s) &\equiv& \int_0^\infty d\tau P(\tau) f(\tau) e^{-s\tau}.
\end{eqnarray}
Then we obtain up to the first order of $r$
\begin{equation}
	\sum_{k=1}^\infty \widetilde{P_k}(s) \approx \frac{ \tilde{P}(s) }{ 1 - \tilde{P}(s) } + \frac{r \tilde{Q}(s)^2}{[1-\tilde{P}(s)]^2} + \mathcal{O}(r^2).
	\label{eq:Pk_s_approx}
\end{equation}
The calculation of the higher-order terms of $r$ is straightforward. By taking the inverse Laplace transform of Eq.~\eqref{eq:Pk_s_approx} and plugging it into Eq.~\eqref{eq:Atd}, we finally get the autocorrelation function as a function of $M$ for the arbitrary form of $P(\tau)$, which is denoted by $A_M(t_d)$ hereafter.

\subsection{Exponential IET distribution}\label{subsec:expo}

One can consider the case with exponentially distributed IETs that are correlated with each other. Despite the fact that it is hard to find real-world examples of this case, we study this case because it is a good testbed for our analytical framework. Precisely, we use the following form of $P(\tau)$ with the mean IET $\mu\gg \tau_{\rm min}=1$: 
\begin{equation}
    P(\tau)= \mu^{-1}e^{-\tau/\mu},
    \label{eq:Ptau_expo}
\end{equation}
by which one gets $a=1/4$ in Eq.~\eqref{eq:Mr_ratio}, hence $r=4M$. From Eq.~\eqref{eq:Ptau_expo}, one gets
\begin{equation}
    \tilde{P}(s) = \frac{1}{\mu s + 1},\
    \tilde{Q}(s) = \frac{-\mu s}{(\mu s + 1)(\mu s + 2)}.
    \label{eq:PQs}
\end{equation}
Plugging Eq.~\eqref{eq:PQs} into Eq.~\eqref{eq:Pk_s_approx} as well as using Eq.~\eqref{eq:Atd}, we analytically derive the autocorrelation function up to the first order of $M$ as
\begin{equation}
    \label{eq:Atd_expo}
    A_M(t_d)\approx \frac{4M t_d e^{-2t_d/\mu}}{\mu(\mu-1)} +\mathcal{O}(M^2),
\end{equation}
where $\lambda=1/\mu$ has been used. Note that the first term on the right hand side in Eq.~\eqref{eq:Atd_expo} can be written as $[4M/(\mu-1)]g(t_d/\mu)$ with $g(x)=xe^{-2x}$, implying that $A_M(t_d)$ for various values of $\mu$ and $M$ can be collapsed when rescaled properly.

\begin{figure}[!t]
  \includegraphics[width=.7\columnwidth]{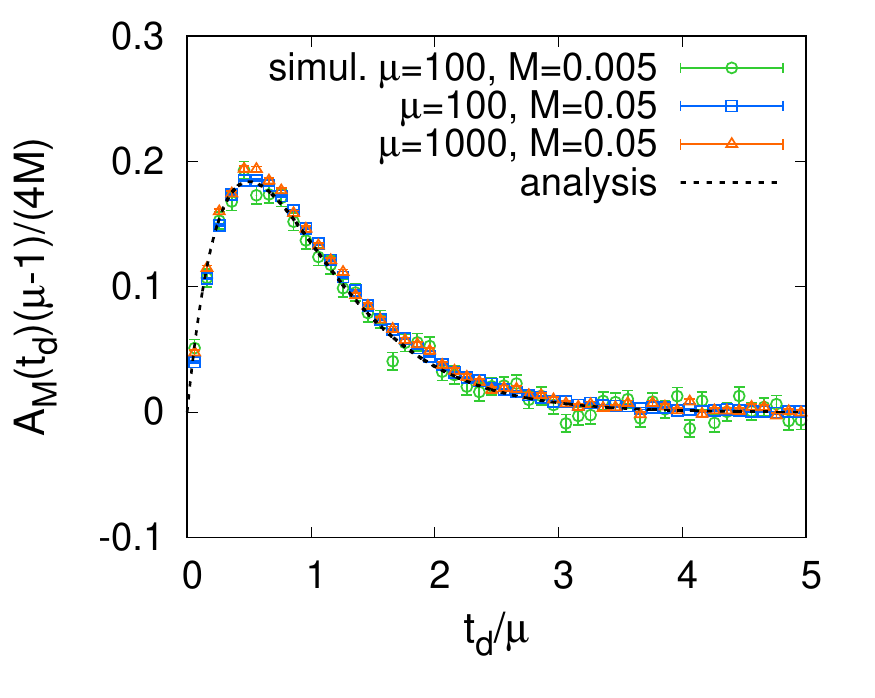}
  \caption{Case with the exponential IET distribution in Eq.~\eqref{eq:Ptau_expo}: Simulation results of the autocorrelation function $A_M(t_d)$ for various values of $\mu$ and $M$ (symbols) are collapsed when rescaled properly. They are in good agreement to the analytical result up to the first order of $M$ in Eq.~\eqref{eq:Atd_expo} (black dotted curve). Each point and its standard error were obtained over $10^4$ event sequences of $n=5\times 10^4$.}
  \label{fig:Ptau_expo}
\end{figure}

For the numerical validation of our analytical result, we introduce an algorithm for generating the event sequence using the FGM copula provided that $P(\tau)$ and $M$ are given, which is called the copula-based algorithm~\cite{Jo2019Copulabased}: To generate a sequence of $n$ IETs, i.e., $\{\tau_1,\cdots,\tau_n\}$, the first IET $\tau_1$ is drawn from $P(\tau)$ and the second IET $\tau_2$ is drawn from the conditional PDF $P(\tau_2|\tau_1)=P(\tau_1,\tau_2)/P(\tau_1)$, where $P(\tau_1,\tau_2)$ is modeled by the FGM copula in Eq.~\eqref{eq:Ptautau}. Then $\tau_i$ for $i=3,\cdots,n$ are sequentially drawn. Once the sequence of $n$ IETs is ready, the timings of $n+1$ events are set to be $t_0=0$ and $t_i=\sum_{i'=1}^i \tau_{i'}$ for $i=1,\cdots,n$; the event sequence $x(t)$ has the value of $1$ for $t\in\{t_0,\cdots, t_n\}$, otherwise $x(t)=0$. This $x(t)$ is then used to calculate the autocorrelation function in Eq.~\eqref{eq:auto_define}.

As shown in Fig.~\ref{fig:Ptau_expo}, the simulation results using Eq.~\eqref{eq:Ptau_expo} for various values of $\mu$ and $M$ are in good agreement to our analytical result in Eq.~\eqref{eq:Atd_expo}.

\subsection{Power-law IET distribution}\label{subsec:power}

To be more realistic, we consider a power-law IET distribution with an exponential cutoff:
\begin{equation}
    P(\tau) = \frac{\tau_{c}^{\alpha-1}}{\Gamma \left( 1-\alpha,1/\tau_{c} \right)} \tau^{-\alpha} e^{-\tau/\tau_{c}}\theta(\tau-1),
    \label{eq:Ptau_cutoff}    
\end{equation}
where $\alpha$ and $\tau_c$ denote the power-law exponent and exponential cutoff, respectively. $\Gamma(\cdot,\cdot)$ is an upper incomplete Gamma function and $\theta(\cdot)$ is a Heaviside step function, implying $\tau_{\rm min}=1$. We also set $\tau_c=10^6$ for the rest of the paper, which is sufficiently large for studying the scaling behavior of the autocorrelation function. With this setup we numerically obtain the value of $a$ in Eq.~\eqref{eq:Mr_ratio}, e.g., $a\approx 0.0039$ for $\alpha=1.4$ and $a\approx 0.0033$ for $\alpha=2.7$, respectively. 

Since the analysis of the autocorrelation function with Eq.~\eqref{eq:Ptau_cutoff} is not straightforward, we instead use a simple power-law function for the IET distribution as
\begin{equation}
    P(\tau)= (\alpha-1) \tau^{-\alpha}\theta(\tau-1),
    \label{eq:Ptau_power}
\end{equation}
which yet allows us to study the scaling behavior of the autocorrelation function to some extent. From Eq.~\eqref{eq:Ptau_power} one gets 
\begin{eqnarray}
    \label{eq:Ps_exact}
    \tilde{P}(s) &=& (\alpha-1)s^{\alpha-1} \Gamma(1-\alpha,s), \\
    \label{eq:Qs_exact}
    \tilde{Q}(s) &=& \tilde{P}(s)-2(\alpha-1)s^{2\alpha-2} \Gamma(2-2\alpha,s).
\end{eqnarray}
We first analyze the case with $1<\alpha<2$. In the asymptotic limit of $s\to 0$ one obtains
\begin{eqnarray}
    \tilde{P}(s) &\approx& 1 + b_1 s^{\alpha-1} +c_1s 
    + \mathcal{O}(s^2), \\
    \tilde{Q}(s) &\approx& b_1 s^{\alpha-1} - b_2 s^{2\alpha-2} +(c_1 -c_2)s + \mathcal{O}(s^2),
\end{eqnarray}
where for $\alpha\neq 3/2$
\begin{eqnarray}
    &&b_1 \equiv \Gamma(1-\alpha)(\alpha-1),\ 
    c_1 \equiv \frac{\alpha-1}{2-\alpha},\nonumber\\
    &&b_2 \equiv \Gamma(2-2\alpha)(2\alpha-2),\ 
    c_2 \equiv \frac{2\alpha-2}{3-2\alpha}.\nonumber
\end{eqnarray}
From Eqs.~\eqref{eq:Atd} and~\eqref{eq:Pk_s_approx}, and with $\lambda=0$ due to the diverging $\mu$, we get for $\alpha\neq 3/2$
\begin{eqnarray}
    && A_M(t_d) \approx B_1 t_d^{-(2-\alpha)} + B_2 t_d^{-(4-2\alpha)} +\cdots \nonumber\\
    && +\frac{M}{a}( C_1 t_d^{-\alpha} + C_2 t_d^{-(3-\alpha)} +\cdots) +\mathcal{O}(M^2), 
    \label{eq:Atd_approx_alpha12}
\end{eqnarray}
where 
\begin{eqnarray}
    && B_1 \equiv \frac{-1}{b_1\Gamma(\alpha-1)},\ 
    B_2 \equiv \frac{c_1}{b_1^2 \Gamma(2\alpha-3)},\nonumber\\
    && C_1 \equiv \frac{-2b_2}{b_1 \Gamma(1-\alpha)},\
    C_2 \equiv \frac{-2c_2}{b_1 \Gamma(\alpha-2)}. \nonumber
\end{eqnarray}
In the case with uncorrelated IETs, i.e., $M=0$, the leading term of $t_d^{-(2-\alpha)}$ leads to the well-known scaling relation of $\alpha+\gamma=2$ for $1<\alpha<2$ in Eq.~\eqref{eq:alpha_gamma}. 

\begin{figure}[!t]
    \includegraphics[width=\columnwidth]{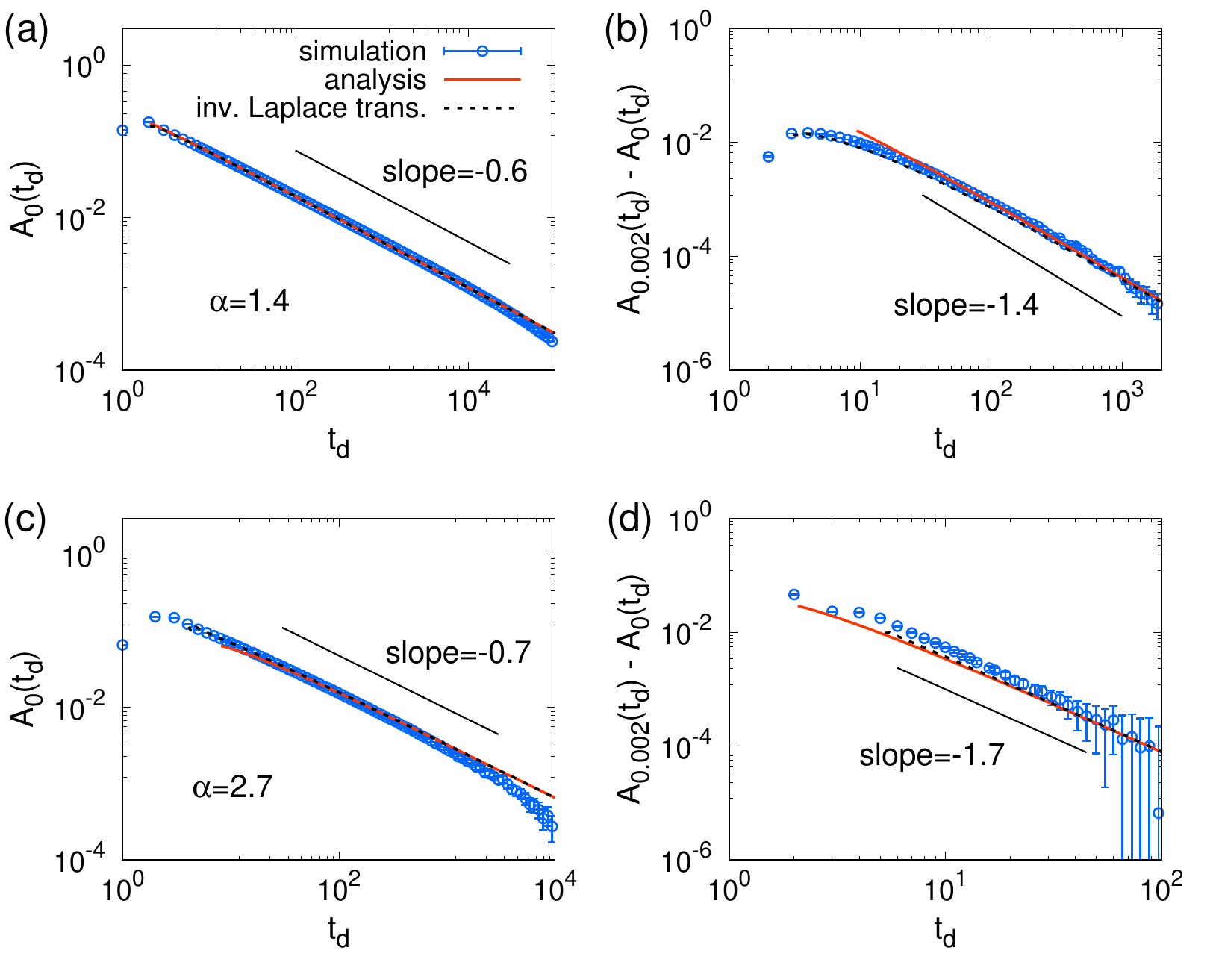}
    \caption{Case with the power-law IET distribution in Eq.~\eqref{eq:Ptau_cutoff} for the simulation [Eq.~\eqref{eq:Ptau_power} for the analysis]: Simulation results of the autocorrelation function $A_0(t_d)$ (a,c) and the difference between $A_{0.002}(t_d)$ and $A_0(t_d)$ (b,d) for $\alpha=1.4$ (a,b) and $2.7$ (c,d) (blue circles) are compared to the corresponding analytical result in Eqs.~\eqref{eq:Atd_approx_alpha12} or~\eqref{eq:Atd_approx_alpha23} (red solid curve). We also plot the curves of $A_0(t_d)$ and $A_{0.002}(t_d)-A_0(t_d)$ that are obtained by the numerical inverse Laplace transform of Eq.~\eqref{eq:Pk_s_approx} with $\tilde{P}(s)$ in Eq.~\eqref{eq:Ps_exact} and $\tilde{Q}(s)$ in Eq.~\eqref{eq:Qs_exact} (black dotted curve). Each point and its standard error were obtained over up to $10^4$ event sequences of $n=5\times 10^4$.}
    \label{fig:Ptau_power}
\end{figure}

The above analytical result in Eq.~\eqref{eq:Atd_approx_alpha12} is to be validated by the simulation results using Eq.~\eqref{eq:Ptau_cutoff}. For the uncorrelated IETs, $A_0(t_d)$ for $\alpha=1.4$ is calculated from the event sequences generated using the copula-based algorithm, as depicted in Fig.~\ref{fig:Ptau_power}(a). The simulation result of $A_0(t_d)$ turns out to be in good agreement with our analytical result in Eq.~\eqref{eq:Atd_approx_alpha12} with $M=0$ for several decades of $t_d$. To confirm the effects due to the correlations between IETs, $A_M(t_d)$ is numerically obtained for $\alpha=1.4$ and $M=0.002$ (i.e., $r=M/a \approx 0.52$). Then we calculate its difference from the uncorrelated case, i.e., $A_{0.002}(t_d)-A_0(t_d)$, which is found to be comparable to the analytical result up to the first order of $M$ in Eq.~\eqref{eq:Atd_approx_alpha12}, see Fig.~\ref{fig:Ptau_power}(b).

Next, we analyze the case with $2<\alpha<3$, where $\mu$ is finite and $\lambda=1/\mu=-1/c_1$, to obtain for $\alpha\neq 5/2$
\begin{eqnarray}
    &&  A_M(t_d) \approx B'_1 t_d^{-(\alpha-2)} + B'_2 t_d^{-(2\alpha-4)} +\cdots \nonumber\\
    && +\frac{M}{a}( C'_1 t_d^{-(\alpha-1)} + C'_2 t_d^{-(2\alpha-3)} +\cdots) +\mathcal{O}(M^2), 
    \label{eq:Atd_approx_alpha23}
\end{eqnarray}
where 
\begin{eqnarray}
    && B'_1 \equiv \frac{b_1}{c_1(c_1+1)\Gamma(3-\alpha)},\
    B'_2 \equiv \frac{-b_1^2}{c_1^2(c_1+1) \Gamma(5-2\alpha)},\nonumber\\
    && C'_1 \equiv \frac{2(c_1-c_2)c_2 b_1}{c_1^2(c_1+1) \Gamma(2-\alpha)},\
    C'_2 \equiv \frac{(3c_2-2c_1)c_2b_1^2}{c_1^3(c_1+1) \Gamma(4-2\alpha)}. \nonumber
\end{eqnarray}
For the case with $M=0$, the leading term of $t_d^{-(\alpha-2)}$ leads to the well-known scaling relation of $\alpha-\gamma=2$ for $2<\alpha<3$ in Eq.~\eqref{eq:alpha_gamma}. 

We find that the simulation results of $A_0(t_d)$ and of the difference of $A_{0.002}(t_d)-A_0(t_d)$ for $\alpha=2.7$ from the event sequences generated using the copula-based algorithm are comparable to our analytical result in Eq.~\eqref{eq:Atd_approx_alpha23}, as evidenced in Fig.~\ref{fig:Ptau_power}(c,d). Note that $M=0.002$ means $r=M/a\approx 0.60$. The discrepancy for the difference of $A_{0.002}(t_d)-A_0(t_d)$ between the analytical and simulation results might be attributed to the finite $\tau_c$ and/or $n$.

\begin{figure}[!t]
    \includegraphics[width=\columnwidth]{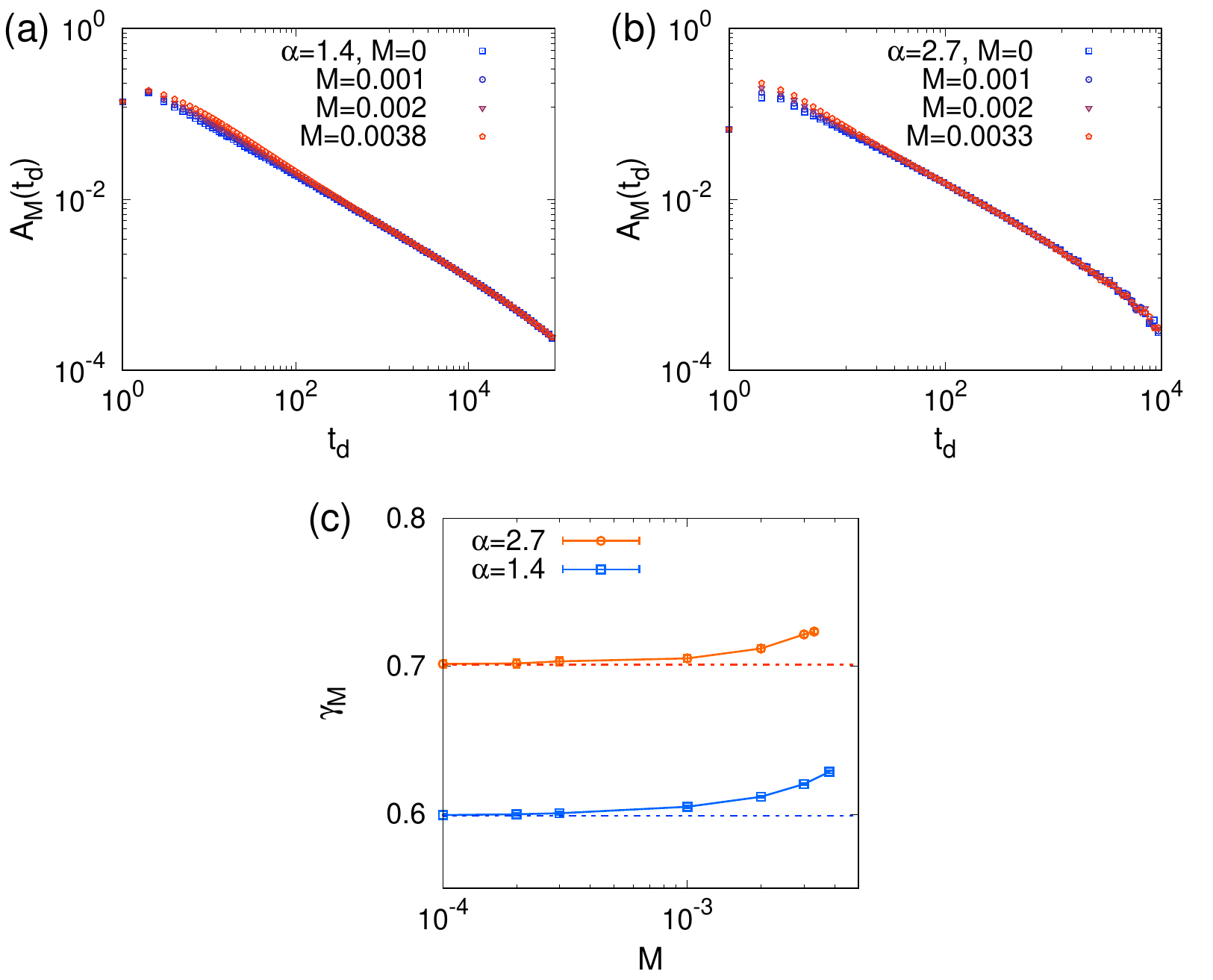}
    \caption{Case with the power-law IET distribution in Eq.~\eqref{eq:Ptau_cutoff}: (a,b) Simulation results of the autocorrelation function $A_M(t_d)$ for various values of $M$ when $\alpha=1.4$ (a) and $2.7$ (b), respectively. The standard errors for the points are smaller than the symbol size. (c) Estimated values of the apparent decaying exponent $\gamma_M$, defined by Eq.~\eqref{eq:simple_gamma}, for several values of $M$. Each point and its standard error were obtained from up to $10^4$ event sequences of $n=5\times 10^4$. For each $\alpha$, the estimated value of $\gamma_0$ for $M=0$ is also plotted by a horizontal dotted line for comparison.}
    \label{fig:gamma_M}
\end{figure}

Finally, we discuss the effect of $M$ on the overall decaying behavior of $A_M(t_d)$. We make two observations in Eq.~\eqref{eq:Atd_approx_alpha12}: (i) The leading term coupled with $M$ is either of the order of $t_d^{-\alpha}$ for $1<\alpha< 3/2$ or of the order of $t_d^{-(3-\alpha)}$ for $3/2<\alpha<2$, and the coefficient of this $M$-coupled leading term is positive, i.e., $C_1>0$ for $1<\alpha<3/2$ and $C_2>0$ for $3/2<\alpha<2$. This indicates that $A_M(t_d)$ for $M>0$ begins with a larger value than that of $A_0(t_d)$ for small $t_d$. (ii) Such $M$-coupled leading term, $t_d^{-\alpha}$ or $t_d^{-(3-\alpha)}$, decays faster than the leading term for $M=0$, which is of the order of $t_d^{-(2-\alpha)}$. This implies that $A_M(t_d)$ for $M>0$ eventually approaches $A_0(t_d)$ for sufficiently large $t_d$. Combining these two observations, we conclude that the stronger correlation between IETs with the larger $M$ results in the steeper decay of $A_M(t_d)$, despite the fact that $A_M(t_d)$ for $M>0$ is always larger than $A_0(t_d)$. This analytical expectation is consistent with the simulation results as depicted in Fig.~\ref{fig:gamma_M}(a). We also observe the similar behavior in Eq.~\eqref{eq:Atd_approx_alpha23} such that the $M$-coupled leading term of $t_d^{-(\alpha-1)}$ has the positive coefficient ($C'_1>0$) and decays faster than the leading term for $M=0$ of the order of $t_d^{-(\alpha-2)}$. The tendency of the steeper decay for the larger $M$ is evident in the simulation results, see Fig.~\ref{fig:gamma_M}(b). Therefore, if the value of decaying exponent is naively estimated using the simple scaling form as 
\begin{equation}
    A_M(t_d)\sim t_d^{-\gamma_M},
    \label{eq:simple_gamma}
\end{equation}
one may find an increasing tendency of the apparent decaying exponent $\gamma_M$ with $M$. This tendency is numerically confirmed for both cases with $1<\alpha<2$ and $2<\alpha<3$, as shown in Fig.~\ref{fig:gamma_M}(c). It is remarkable from both analytical and simulation results that even a little amount of the correlation between IETs can change the apparent decaying exponent $\gamma_M$, implying that the scaling relations in Eq.~\eqref{eq:alpha_gamma} can be easily violated by the correlations between IETs.

\section{Conclusion}\label{sec:conclusion}

In order to investigate the effects of correlations between interevent times (IETs) on the autocorrelation function, we have derived the analytical form of the autocorrelation function for the arbitrary IET distribution $P(\tau)$ and for small values of the memory coefficient $M$, i.e., in the case with weakly correlated IETs, where the Farlie-Gumbel-Morgenstern copula~\cite{Nelsen2006Introduction, Takeuchi2010Constructing} is adopted for modeling the joint probability distribution function of two consecutive IETs. For the numerical validation, the event sequences are generated using the copula-based algorithm~\cite{Jo2019Copulabased}, by which IETs can be drawn sequentially only conditioned by their previous IETs. For both exponential and power-law IET distributions, we find that the simulation results of autocorrelation functions are in good agreement with the corresponding analytical solutions. 

In particular, for the power-law case, we find that the stronger correlation between IETs with the larger $M$ leads to the steeper decay of the autocorrelation function. In other words, the apparent decaying exponent $\gamma$ is found to increase with $M$. Our finding sheds light on the effects of correlations between IETs on other measures for temporal correlations too, such as Hurst exponent $H$ and the scaling exponent of the power spectral density $\eta$, considering their interdependence~\cite{Kantelhardt2001Detecting, Allegrini2009Spontaneous, Rybski2009Scaling, Rybski2012Communication}. We also expect to better understand the differences between the empirical autocorrelation functions and those calculated for the randomized event sequences~\cite{Karsai2012Universal, Rybski2012Communication} based on our results. Finally, our results also support the previous numerical finding on the increasing tendency of $\gamma$ for the stronger correlation between IETs~\cite{Jo2017Modeling}, where the correlations between IETs have been controlled by the power-law exponent of bursty train size distributions. Here we like to note that the bursty train size distribution and $M$ have been related to each other~\cite{Jo2018Limits}.

We remark that our analytical approach has limits as follows: (i) We have considered only the correlations between two consecutive IETs based on the empirical findings, while the correlations between the arbitrary number of consecutive IETs have also been empirically observed in terms of heavy-tailed distributions of bursty train sizes~\cite{Karsai2012Universal, Yasseri2012Dynamics, Wang2015Temporal}. This requires us to devise the more general analytical approach than ours as a future work. (ii) The FGM copula allows only relatively weak correlations between IETs, requiring us to consider other copulas for the cases with the stronger correlation between IETs~\cite{Nelsen2006Introduction}. Despite such limits, our analytical approach can help us to better understand the long-term temporal correlations ubiquitously observed in various natural and social phenomena, as yet little is known about the effects of the correlations between IETs on the long-term temporal correlations.

\begin{acknowledgments}
    The author thanks Takayuki Hiraoka for fruitful discussions and acknowledges financial support by Basic Science Research Program through the National Research Foundation of Korea (NRF) grant funded by the Ministry of Education (NRF-2018R1D1A1A09081919).
\end{acknowledgments}

\bibliographystyle{apsrev4-1}
%

\end{document}